\documentclass[preprint,nofootinbib,superscriptaddress] {revtex4}

\usepackage{amsmath, amssymb, bm, mathrsfs}
\usepackage{graphicx}
\usepackage{color}
\usepackage{times}
\usepackage[english]{babel}
\usepackage[latin1]{inputenc}

\newcommand {\bfr} {\bm{r}} 
\newcommand {\dd} {{\mathrm d}} 
\newcommand {\half} {\frac{1}{2}}
\newcommand{\del}{\partial}
\newcommand{\kB}{k_{\mathrm B}} 
\newcommand{\Eqref}[1]{Eq.~\eqref{#1}}

\newcommand{\LL}{\bm{\Lambda}} 
\newcommand{\GG}{\bm{\Gamma}} 
\newcommand{\DD}{D_0} 
\newcommand{\MM}{m} 
\newcommand{\rot}{\mathcal{R}} 
\newcommand{\rotb}{\bm{\rot}} 
\newcommand{\te}{{t_\mathrm{f}}}
\newcommand{\mT}{\intercal} 
\def\bra#1{\mathinner{\langle{#1}|}}
\def\ket#1{\mathinner{|{#1}\rangle}}
\newcommand{\braket}[2]{\langle #1|#2\rangle}

\makeatletter \def\@dotsep{4.5} \makeatother
\newcommand{\im}{{\mathrm i}}
\newcommand{\ee}{{\mathrm e}}
\newcommand{\ie}{{i.e}}

\newcommand{\eg}{{e.g}}
\newcommand{\etc}{{etc}}

\begin{document}

\title{NMR signal for particles diffusing under potentials: From path integrals and numerical methods to a new model of diffusion anisotropy}

\author{Cem Yolcu}\affiliation{Department of Physics, Bo\u gazi\c ci University, Bebek, \. Istanbul, 34342, Turkey}

\author{Muhammet Memi\c c}\affiliation{Department of Physics, Bo\u gazi\c ci University, Bebek, \. Istanbul, 34342, Turkey}

\author{Kadir \c Sim\c sek}\affiliation{Department of Physics, Bo\u gazi\c ci University, Bebek, \. Istanbul, 34342, Turkey}

\author{Carl-Fredrik Westin}
\affiliation{Department of Radiology, Brigham and Women's Hospital, Harvard Medical School, Boston, MA, 02215, USA}

\author{Evren \"Ozarslan}
\email[Electronic address: ]{evren.ozarslan@boun.edu.tr}
\affiliation{Department of Physics, Bo\u gazi\c ci University, Bebek, \. Istanbul, 34342, Turkey}
\affiliation{Department of Radiology, Brigham and Women's Hospital, Harvard Medical School, Boston, MA, 02215, USA}


\begin{abstract}
We study the influence of diffusion on NMR experiments when the molecules undergo random motion under the influence of a force field, and place special emphasis on parabolic (Hookean) potentials. To this end, the problem is studied using path integral methods. Explicit relationships are derived for commonly employed gradient waveforms involving pulsed and oscillating gradients. The Bloch-Torrey equation, describing the temporal evolution of magnetization, is modified by incorporating potentials. A general solution to this equation is obtained for the case of parabolic potential by adopting the multiple correlation function (MCF) formalism, which has been used in the past to quantify the effects of restricted diffusion. Both analytical and MCF results were found to be in agreement with random walk simulations. A multi-dimensional formulation of the problem is introduced that leads to a new characterization of diffusion anisotropy. Unlike for the case of traditional methods that employ a diffusion tensor, anisotropy originates from the tensorial force constant, and bulk diffusivity is retained in the formulation. Our findings suggest that some features of the NMR signal that have traditionally been attributed to restricted diffusion are accommodated by the Hookean model. Under certain conditions, the formalism can be envisioned to provide a viable approximation to the mathematically more challenging restricted diffusion problems.
\end{abstract}



\maketitle

\section {Introduction}

Through its influence on the diffusion of spin-bearing particles, the physical characteristics of the medium leaves its signature on the observed NMR signal. This makes NMR a powerful probe into the microstructure of biological specimens as well as other porous media. NMR's sensitivity to diffusional processes can be controlled by introducing gradient waveforms into standard NMR experiments \cite{StejskalTanner65}.

The effect of free diffusion on experiments with general gradient waveforms has been fully characterized \cite{Stejskal65,Karlicek80}. However, the complexity of the environment typically leads to non-Gaussian motion, which yields interesting features in the NMR signal decay that depend on the local structure. For example, a time-dependence in the diffusion coefficients \cite{Latour94,Sen04review,Ozarslan_JMR06} tend to emerge in complex environments, while the large wavenumber regime of the signal decay could exhibit unique features \cite{Callaghan91,Ozarslan_JMR07} as well. Time dependence in diffusion coefficients was investigated extensively with emphasis on \hbox{(non-)ergodicity} and ageing \cite{Bodrova15pccp, Bodrova15njp, Cherstvy15}. It has also been studied as a possible indicator of mesoscopic disorder in materials and tissues, making use of diffusion-sensitive NMR images \cite{Novikov11, Novikov14} (see Ref.\ \cite{Sykova08review} for a comprehensive review on diffusion in tissues). Fractional Brownian motion is yet another viable model in which to view the signatures of non-Gaussian motion in NMR signals, which is amenable to path integral methods \cite{Fan15} similar in that regard to the approach discussed in the present article.

In this paper, we consider the case of diffusion taking place under the influence of a Hookean force field. This problem is well-studied in the field of stochastic processes \cite{Smoluchowski13,Smoluchowski16,Uhlenbeck30}. To our knowledge, the first occurrence of it in the NMR literature is in a couple of papers published in 1960s \cite{Stejskal65,Tanner68}, wherein the authors consider the case of ``diffusion near an attractive center'' and derive the signal expression for their then recent pulsed field gradient framework \cite{StejskalTanner65}. Since then Callaghan and Pinder used the formalism on a semi-dilute solution of an entangled polystyrene to model the displacements of the resulting network, which is stable during the timescale of the NMR measurement \cite{Callaghan80,Callaghan84}. Le Doussal and Sen employed the problem as an ``artificial pore'' model to understand the effects of restricted diffusion in the presence of nonlinear magnetic field gradients \cite{LeDoussal92}. Mitra and Halperin also touched upon the problem, referring to it as the ``parabolic pore'' in the context of the center-of-mass propagators \cite{MitraHalperin95}. 

In this article, we have two main goals: (i) to provide an analytical framework with which one can derive explicit relationships for the signal decay for a general time-dependent gradient waveform, and (ii) develop semi-analytical and numerical tools, which could be generalized with relative ease to complicated potentials. As demonstrated below, our formulation naturally leads to a new characterization of diffusion anisotropy; similar in spirit to the diffusion tensor model \cite{Basser94b}. However, while the latter tries to fit imaging data to a free diffusion model despite the underlying non-free diffusion, the present framework allows taking confinement into account, which is modeled as a harmonic force. The advantage of taking confinement into account is found in the time dependence of the predicted signal profile, which exhibits features more similar to that of actual restricted diffusion. Employing a harmonic confinement instead of a direct implementation of restricted diffusion, on the other hand, offers a degree of tractability that encodes full anisotropy easily. For instance, compared to a restricted diffusion model involving a capped cylinder \cite{Ozarslan_JMR09}, which has \emph{two} length parameters due to its partial isotropy, the confinement tensor generally has three distinct size parameters (related to its eigenvalues) that can capture full three-dimensional anisotropy, much like the diffusion tensor model.

The article is organized as follows: In the next section, Sec.\ \ref{sec:pathint}, we express the NMR signal as a path integral and evaluate it assuming a Hookean potential as stated above, providing a general expression that can be used with arbitrary gradient sequences. As such, we end the section by deriving analytical expressions for the expected NMR signal for the conventional pulsed field gradient (Stejskal-Tanner) and oscillating gradient sequences.  For the sake of keeping the main text to the point, technicalities involved in evaluating the path integral are discussed in great detail in Appendix \ref{app:pathint}. In the subsequent section, Sec.\ \ref{sec:MCF}, we introduce the formulation of the semi-analytical multiple correlation function framework (MCF) \cite{Robertson66,Barzykin98,Grebenkov07,OzarslanJCP09}, assuming a general confining potential. Its application to the special case of a Hookean potential is presented in Appendix \ref{app:MCF}, along with relevant technical details. In Sec.\ \ref{sec:sim}, the random walk simulations are described before proceeding with the validation of the (semi-)analytical approaches and the deliberation of the Hookean model as an alternative tool for the characterization of diffusion anisotropy in the Results section, followed by concluding remarks.

\section {Analytical results through a path integral formalism} 
\label{sec:pathint}

\subsection{General gradient waveforms}

Here, we derive an analytical expression via a path integral for the
average transverse magnetization in a specimen, stemming from
spin-bearing random walkers subject to a Hookean potential. In order
to isolate details of the derivation from the results, we have placed
most of the former in Appendix \ref{app:pathint} while restricting to
the essential points in this section.

In a medium subjected to a (spatially) non-uniform magnetic field, a
diffusing spin-carrier experiences a rate of (Larmor) precession that
varies with respect to its spatial position. Compared to a steady
precession in the case of a uniform magnetic field, therefore, each
particle picks up a phase in its rotation, depending on its random
path. Hence, from an ensemble of such random walkers in a
(time-dependent) spatially linear magnetic field gradient $\bm{G} (t)$
of duration $\te$, an average signal of
\begin{align}
E = \left \langle \exp \! \left( - \im \gamma
\int_0^{\te} \dd t \, \bm{G} (t) \cdot \bfr (t) \right) \right \rangle
\;, \label{eq:E.expectn}
\end{align}
is obtained, which is the average transverse magnetization up to a
unit. Here, since the average is to be taken over all random paths
$\bfr (t)$, the expectation value has the form of a path integral,
\begin{align}
E  = \int \mathscr{D} \bfr (t) \mathscr{P} [\bfr (t)] \,
\ee^{- \im \gamma \int_0^{\te} \dd t \, \bm{G} (t) \cdot \bfr (t)} \;,
\label{eq:path.int}
\end{align}
where $\int \mathscr{D} \bfr (t)$ denotes an integral over the space
of paths, and $\mathscr{P} [\bfr (t)]$ the weight of the path $\bfr
(t)$. Note that $E$ is a normalized quantity that takes the value of 1
when $\bm G = 0$, and $\gamma$ is the gyromagnetic ratio. The
probability weight $\mathscr{P} [\bfr (t)]$ is most easily imagined as
an infinite product of stepwise probabilities of transition from each
intermediate path point to the next.

In this article, we assume that the stochastic process underlying the random paths is diffusion with diffusivity $\DD$ under the effect of a \emph{dimensionless} Hookean potential,\footnote{In this article, we customarily use boldface symbols for \emph{collections} of numbers, be them vectors or matrices.}
\begin{align}
  V (\bfr) = \frac 12 \bfr^\mT \bm{C} \bfr \ ,
\end{align}
where we refer to $\bm{C}$, having dimension of inverse length squared, as the \emph{confinement tensor}. The true tensorial force constant can be obtained by multiplying $\bm C$ by the Boltzmann constant $\kB$ and the absolute temperature $T$. 

We argue that the potential defined above suffices to capture relevant features of diffusion in confined spaces. The advantage of the assumption is that the stepwise probabilities mentioned above have a simple form (see Appendix \ref{app:pathint}) and the resulting path weight is analytically tractable. We find that the NMR signal, or average transverse magnetization, that ensues the application of an arbitrary gradient waveform $\bm{G} (t)$ is expressed as
\begin{align}
E = \exp \! \left( - \DD \int_0^\te \dd t \, | {\bm{Q}} (t)|^2 - \frac
{\DD}{2} {\bm{Q}}^\mT (0) \bm{\Omega}^{-1} {\bm{Q}} (0) \right)
, \label{eq:E.3D}
\end{align}
where
\begin{align}
\bm{Q} (t) = \gamma \int_t^\te \dd t' \, \ee^{-\bm{\Omega} (t'-t)} 
\bm{G} (t') \ , \label{eq:Q.3D}
\end{align}
and 
\begin{align}
\bm{\Omega} = \DD \bm{C}
\end{align}
is a matrix of rate constants (inverse time), simply proportional to
the confinement tensor $\bm{C}$. The inverse of this matrix encodes the
time scales involved in equilibration, namely the approach of a
distribution of diffusing particles toward the Boltzmann distribution
$\sim \ee^{-V (\bfr)}$. Essentially, it is the finite width of
this final Gaussian profile that is regarded as a measure of
confinement in the present article, by matching it to the width of an
actual confined geometry (Appendix \ref{sec:analytical1D}).

\subsection{Explicit results for common gradient waveforms} 
\label{sec:ExplicitResults}

\begin{figure}
  \begin{center}
    \includegraphics[width=8.4cm]{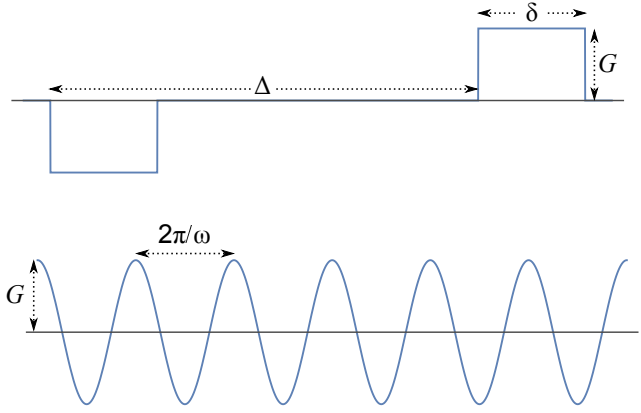}
    \caption{(Color online) Top: Stejskal and Tanner's pulsed field gradient
      sequence. Bottom: Oscillating gradient
      waveform.}  \label{fig:waveforms}
  \end{center}
\end{figure} 

The results derived above can be employed to obtain exact expressions
for the NMR signal intensity for any gradient waveform. In this
section, we shall consider two widely employed experiments illustrated
in Figure \ref{fig:waveforms}. The first sequence is Stejskal and
Tanner's pulsed field gradient technique \cite{StejskalTanner65} while
the second one features an oscillatory (sinusoidal) profile
\cite{Stepisnik81}.

\subsubsection{Stejskal-Tanner sequence}

Here, we use Eqs.\ \eqref{eq:E.3D} and \eqref{eq:Q.3D} to predict the
NMR signal that should be recovered from spin-bearing random walkers
subject to a three dimensional harmonic potential in a traditional
pulsed field gradient experiment introduced by Stejskal and Tanner
\cite{StejskalTanner65}. As illustrated on the top panel in Figure
\ref{fig:waveforms}, the effective waveform comprises two gradient
pulses in opposite directions with a delay $\Delta$ between their
leading edges. Each pulse has magnitude $G$ and duration $\delta$. Due
to the tractability of our ``harmonic approximation,'' an expression
for the signal can be obtained easily, valid for all possible values
of the experimental parameters.

The first thing to do is to evaluate the integral~\eqref{eq:Q.3D} for $\bm{Q} (t)$, which is straightforward (as noted before, one may wish to consider temporarily the eigenbasis). One finds (taking the transpose for convenience), 
\begin{widetext}
	\begin{align}
	\renewcommand\arraystretch{1.33}
	\bm{Q}^\mT (t) = \gamma \bm{G}^\mT \bm{\Omega}^{-1} 
	\left \{ \begin{array} {lll}
	\bm{1} - \left[\ee^{-\bm{\Omega} \Delta} + 
	\ee^{-\bm{\Omega} \delta} - \ee^{-\bm{\Omega} (\Delta + \delta)} 
	\right] \ee^{\bm{\Omega} t}  
	&, & 0 < t < \delta \\
	\left[ \ee^{-\bm{\Omega} (\Delta+\delta)} - \ee^{-\bm{\Omega} \Delta} 
	\right] \ee^{\bm{\Omega} t} 
	&, & \delta < t < \Delta \\
	\ee^{-\bm{\Omega}(\Delta+\delta)} \ee^{\bm{\Omega} t} 
	- \bm{1} 
	&, & \Delta < t <\Delta +\delta \\
	0 &, & \Delta + \delta < t \end{array} \right . \;,\label{eq:Q.1PFG}
	\end{align}
	where $\bm{1}$ represents the $3\times 3$ identity
        matrix. What remains is the tedious task of
        squaring\footnote{By the square of the vector $\bm{Q}$ we of
          course mean $\bm{Q}^\mT \bm{Q}$. It is useful to note that
          all the matrices that will end up between $\bm{Q}^\mT$ and
          $\bm{Q}$ according to Eqs.\ \eqref{eq:E.3D} and
          \eqref{eq:Q.1PFG} commute (since they all are eventually
          powers of $\bm{\Omega}$) and are symmetric, simplifying the
          algebra greatly. \label{ft:algebra}} and integrating this
        expression to evaluate the signal \eqref{eq:E.3D}. After a bit
        of algebra and simplifications, we obtain the quadratic
        expression (in $\bm{G}$)
	\begin{align}
	E = \exp \!\left(- \bm{G}^\mT \bm{\mathcal{A}} \bm{G}
	\right) \;,  \label{eq:E.1PFG}
	\end{align}
	where the real symmetric matrix
	\begin{align}
	\bm{\mathcal{A}} = \DD \gamma^2 \bm{\Omega}^{-3} \Big[ \left( \bm{1} -
	\ee^{-\bm{\Omega} \Delta} \right) \left( \bm{1} - \ee^{-\bm{\Omega}
		\delta} \right)^2 \ee^{\bm{\Omega} \delta} -
	\left(\bm{1} -\ee^{-2\bm{\Omega} \delta} \right) \ee^{\bm{\Omega}
		\delta} + 2 \bm{\Omega} \delta \Big] \ . \label{eq:Amatrix}
	\end{align}
\end{widetext}

The last two expressions provide a model alternative to the diffusion tensor model \cite{Basser94,Basser94b} commonly used in biomedical applications of magnetic resonance imaging. The main difference is the dependence of the predicted signal on the timing parameters of the sequence. In diffusion tensor imaging (DTI), this dependence is the same as that for free (Gaussian) diffusion. In the Hookean potential model, the dependence is consistent with restricted diffusion \cite{OzarslanJCP08}.

As a check, one may evaluate Eqs.\ \eqref{eq:E.1PFG} and \eqref{eq:Amatrix} to order $\delta^3$, to obtain
\begin{align}
\ln E \approx - \DD \gamma^2 \delta^2 \bm{G}^\mT \bm{\Omega}^{-1} 
\left( \bm{1} - \ee^{-\bm{\Omega} \Delta} - \frac {\bm{\Omega} \delta} {3} 
\right) \bm{G} \;,  \label{eq:Stejskal65}
\end{align}
which is the anisotropic generalization of Stejskal's result \cite{Stejskal65} for a
spherically symmetric harmonic potential, with the additional $\mathcal{O} (\delta^3)$ term. In the $\Omega \rightarrow 0$ (\ie., $C \rightarrow 0$) limit, the Stejskal-Tanner expression for free diffusion is recovered \cite{StejskalTanner65} as expected.

\subsubsection{Oscillating gradient}

Next, we shall consider the sinusoidal waveform with angular frequency $\omega$ as depicted in the bottom of Figure \ref{fig:waveforms}.  Owing to the simplicity of diffusion inside a harmonic potential, we were able to derive an analytical expression for the average magnetization for this waveform as well. Since the gradient is applied in a fixed direction,  the formulation in Appendix \ref{sec:analytical1D} is sufficient to obtain the exact result. 

The gradient profile is of the form $G(t) = G \cos (\omega t + \varphi)$. We consider an experiment with $N$ full periods, i.e., $G(t)$ vanishes outside the interval $0 \le t \le 2 \pi N / \omega$. One finds, from \Eqref{eq:Q(t)}, that 
\begin{align}
Q(t) &= \frac {\gamma G} {\sqrt{\Omega^2 + \omega^2}} \left[ \cos
(\omega  t + \varphi_+) - \cos \varphi_+ \ee^{\Omega (t- \frac {2\pi N}{\omega})} \right] \;, 
\end{align}
where $\varphi_+ = \varphi + \cot^{-1}(\Omega / \omega)$. The signal
then follows from \Eqref{eq:E.cont} after some rearrangement as
\begin{align}
  \frac {\Omega^2 + \omega^2} {\DD \gamma^2 G^2} \ln E =
  \frac {\cos\varphi_- \cos\varphi_+} {\Omega}
\left( 1 - \ee^{-2\pi N \frac {\Omega}{\omega}} \right)- \frac {\pi N}
{\omega}  \;,
\end{align}
with $\varphi_- = \varphi - \cot^{-1} (\Omega/ \omega)$.

\section {MCF formalism} \label{sec:MCF}

In this section, we shall visit the same problem using the multiple
correlation function (MCF) formalism, which has been used in the past
to characterize the effect of restricted diffusion on the NMR
signal. Technical details may be found in Appendix \ref{app:MCF}. We
refer the reader to Ref. \cite{Yolcu15Dagstuhl} for a recent review of
the technique consistent with the notation here, and its relations
with the path integral framework discussed in the previous section.


The multiple correlation function formalism aims to compute the NMR signal 
\begin{align}
E (t) = \int \dd \bfr \, \MM (\bfr, t)  \label{eq:E.MCF}
\end{align}
from the time evolution of the (appropriately-normalized)
magnetization density $\MM (\bfr, t)$. In the presence of a potential
$V(\bfr)$, diffusion is governed by the Smoluchowski equation
\cite{Smoluchowski16}, while the transverse magnetization carried by
the random walkers need also be taken into account. This suggests
that, under a magnetic field gradient waveform $\bm{G} (t)$, the
evolution of $\MM (\bfr, t)$ is governed by an equation akin to the
Bloch-Torrey equation \cite{Torrey56}, which we shall refer to as the
Bloch-Torrey-Smoluchowski (BTS) equation,
\begin{align}
\frac {\del} {\del t} \MM (\bfr, t) = & \DD \nabla \cdot \left(
\ee^{- V(\bfr)} \nabla \ee^{V(\bfr)} \MM (\bfr, t)
\right) \nonumber \\ &- \im \gamma \bm{G} (t) \cdot \bfr \, \MM
(\bfr, t) \;, \label{eq:BTS}
\end{align}
where this time the diffusion term deals also with the potential. Clearly, the above expression is reduced to the Bloch-Torrey equation when the potential is zero.


The solution $\MM (\bfr, t)$ of the BTS equation \eqref{eq:BTS} is
best considered in the abstract function space, in terms of a
propagator responsible for the evolution of the magnetization in
time. Due to the explicit time dependence of the operator on the right
hand side of \Eqref{eq:BTS}, the associated propagator is formally a
time-ordered product of operators of the form $\exp \{ \Delta t [\LL +
  \GG (t)] \}$ over successive time intervals of length $\Delta
t$. The abstract operators $\LL$ and $\GG(t)$ correspond, respectively,
to the two operators whose position-space representations appear on
the right hand side of \Eqref{eq:BTS}. We will refer to $\LL$ as the
Smoluchowski operator, which governs the evolution of the spin
population under an arbitrary potential $V(\bfr)$.

In Appendix \ref{app:MCF}, we show that the signal \eqref{eq:E.MCF}
ends up being expressable in the eigenbasis of the Smoluchowski
operator as the matrix element of the propagator corresponding to the
stationary eigenvalue: Denoting the left and right eigenvectors with
$\bra{w_k}$ and $\ket{u_k}$, the signal has the form
\begin{align}
  E = \bra{w_0} \sideset{}{'} \prod_t \ee^{\Delta t [\LL + \GG (t)]} \ket{u_0} \ , \label{eq:E.MCF.2}
\end{align}
where the prime on the product indicates time ordering (i.e., the
earliest factor acts first), and $t$ is to be understood as a time point representative of each time interval. With this perspective, the MCF approach
relies on using the eigenbasis of the Smoluchowski operator to carry
out the matrix operations involved in evaluating the propagator and
thereby the signal \eqref{eq:E.MCF.2}. The two operators $\LL$ and $\GG$ are expressed in
the eigenbasis of $\LL$, which is done analytically in Appendix
\ref{app:MCF} for the case of a Hookean potential. The remaining
operations of exponentiation and matrix product are eventually carried
out by software according to the waveforms we consider in the Results
section.

Note that while a general time dependence requires a continuum limit
($\Delta t \to 0$) in \Eqref{eq:E.MCF.2},
piecewise-constant waveforms such as those of pulsed-field-gradient
experiments do not. The latter situation lends itself readily to a
product of constant exponential operators over the plateaus of the
waveform, without any discretization error. On the other hand, the
case of a general time dependence is treated as a controlled
approximation based on the size of $\Delta t$. Another inevitable
source of numerical error is the need to truncate the matrix
representations, which are in principle infinite. However, one simply
has to include all eigenfunctions whose time scales of decay (i.e., inverses of their eigenvalues) are
significant compared to the time step $\Delta t$.

\section {Random walk simulations} \label{sec:sim}

To provide a validation of the theoretical developments described above, we simulated random walks taking place under a Hookean potential and computed the phase accumulated by each random walker, which was subsequently used in the estimation of the NMR signal. To this end, we performed biased Bernoulli walk simulations using Python. For the purposes of the present article, we have employed these simulations only in one spatial dimension.

Briefly put, each particle's motion is generated based on Bernoulli
trials performed at uniform time intervals $\tau$. The trials are
biased in the sense that the probabilities $p_+=p$ and $p_-=1-p$ of
changing the particle's coordinate by $a$ or $-a$, respectively, are
not equal due to the presence of a force. The desired diffusive
character---corresponding to a bare diffusion constant $\DD$---of a
random walk generated as such is met by the condition 
\begin{align} \label{eq:RWcond1}
\DD = \frac{a^2}{2\tau} \ .
\end{align}
On the other hand, if the drift velocity $(p_+-p_-)a/\tau$ of the random walker is supposed to match the drift velocity ${(\kB T)}^{-1} \DD F$ of a Brownian mover under a force $F$, then the bias must be 
\begin{align} \label{eq:RWcond2}
p_+-p_- = \frac{\DD \tau}{\kB T a}F = \frac{a}{2 \kB T} F \ . 
\end{align}
The conditions in \eqref{eq:RWcond1} and \eqref{eq:RWcond2} fix two of the simulation parameters $p$, $a$, and $\tau$. As a final condition, one may set the resolution of the simulations by choosing $\tau$ to be smaller than the thermalization time scale $\Omega^{-1}=(D_0 C)^{-1}$ of Eq. (\ref{eq:OSS}a) by a desired amount.

Generating random particle trajectories according to the scheme
described above, at each step we can compute the phase that would be
accumulated due to a magnetic field gradient, and hence ``measure''
the NMR signal resulting from a large number of trajectories. The
simulated signal values which will be presented in the next section
have been computed thusly.

\section{Results}

\subsection{Validation of the approach} 

Here, we present and compare results obtained from (i) the application of the analytical expressions of Sec.\ \ref{sec:pathint}, (ii) the numerical implementation of the MCF scheme of Sec.\ \ref{sec:MCF}, and (iii) simulated data as described in Sec.\ \ref{sec:sim}. To this end, we consider the same two waveforms discussed in Sec.\ \ref{sec:ExplicitResults} and illustrated in Figure \ref{fig:waveforms}.

\begin{figure}[!t]
  \begin{center}
    \includegraphics[width=8.4cm]{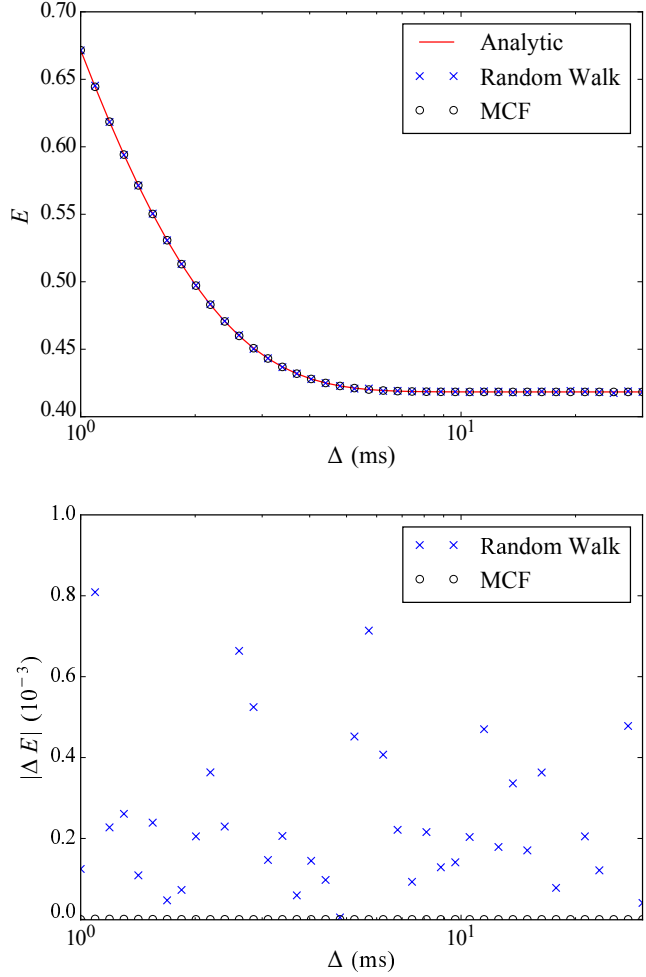}
    \caption{(Color online) Top: The predicted NMR signal plotted
      against the gradient pulse separation (diffusion time) in a
      Stejskal-Tanner pulse sequence. Bottom: Errors incurred in MCF
      and random walk results. The plotted values are the absolute
      values of the deviations from the analytical
      results.}  \label{fig:StejskalTanner_Delta}
  \end{center}
\end{figure}

\paragraph{Stejskal-Tanner sequence.}
The first diagram in Fig.\ \ref{fig:StejskalTanner_Delta} depicts the NMR signal as a function of pulse separation $\Delta$ at fixed $q=\gamma G \delta$. The pulse duration was $\delta = 1\,$ms, and the value of $G$ was chosen so that the wavenumber is $q/2\pi =100\,$mm$^{-1}$.  The temperature and bulk diffusivity were taken to be  $T=310\,$K, and $\DD=3\,\mu$m$^2$/ms, respectively. The force constant was chosen such that $C = 0.33\, \mu$m$^{-2}$;  through Eq.~\eqref{eq:Leff}, this leads to an effective pore size $L_\mathrm{eff} = 6\,\mu$m, which is roughly the size of a red blood cell. The analytical result is due to the application of \Eqref{eq:E.1PFG}, whereas the numerical implementation of the MCF signal \eqref{eq:E.MCF.2} is achieved via Eqs.\ \eqref{eq:signal.exp}--\eqref{eq:Rkl}. The random walk simulations were performed using a step size of $a=0.1\,\mu$m, and a time step of $\tau=a^2/2 \DD = 1.67\, \mu$s. A total of 2 million trajectories were generated.

The decay of the signal with increasing diffusion time is seen to be captured faithfully by the MCF implementation and random walk simulations as well. In Fig.\ \ref{fig:StejskalTanner_Delta}(bottom), we show the absolute error of MCF and the simulations. The MCF implementation is error-free in the PFG scheme, since the pulse sequence \emph{is} actually piecewise constant.

\paragraph{Oscillating gradient waveform.}

\begin{figure}[!t]
  \begin{center}
    \includegraphics[width=8.4cm]{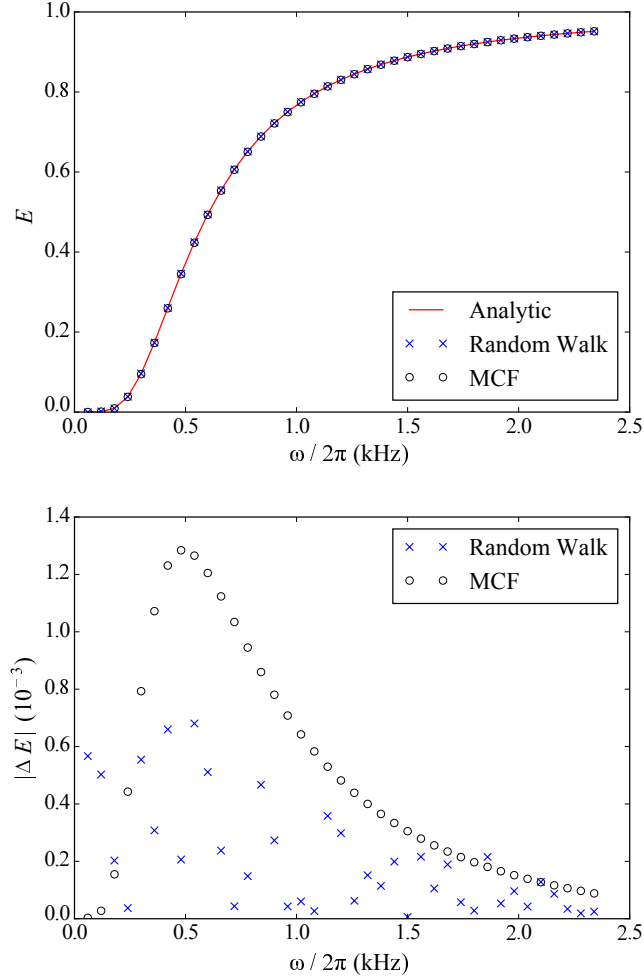}
    \caption{(Color online) Top: The predicted NMR signal plotted against the frequency in an oscillating gradient sequence. Bottom: Errors incurred in MCF and random walk results. The plotted values are the absolute values of the deviations from the analytical results.} \label{fig:OGSE}
  \end{center}
\end{figure}

We show the results for the oscillating waveform in Figure \ref{fig:OGSE}. The same values for the temperature, bulk diffusivity, and the force constant, hence the effective pore size, were used. The phase shift of the waveform was taken to be $\varphi =0$, \ie.,  $G(t)=G \, \cos(\omega t)$ with $G=1\,$T/m.  In these simulations, the total duration of the waveform was fixed at $100\,$ms for all frequencies. Frequency was varied by changing the number of full periods ($N$) within this timeframe. So, $N=1$ corresponds to the lowest frequency, while the subsequent points were generated by setting $N=2,3,\ldots$ The MCF results were obtained by setting the time interval to $10\, \mu$s for all frequencies. The random walk simulations were performed in the same way as for the case of the Stejskal-Tanner sequence.

The frequency dependence of the signal seems to be consistent in all three methods. The errors in the random walk and MCF results are rather small as shown in the bottom panel of Figure \ref{fig:OGSE}. Unlike before though, the MCF results do exhibit some deviation from the analytical results. At the low end of the frequency range, errors of the MCF scheme approach zero, since its staircase approximation of a slowly-varying waveform is sufficiently faithful. As the waveform explores increasingly higher frequencies, the approximation is seen to worsen until the period of the waveform reaches the thermalization timescale $\Omega^{-1} = (D_0 C)^{-1} \simeq 1$ms. Beyond this neighborhood of frequencies, less and less time remains for appreciable diffusion to take place in each period of the gradient waveform. Since the sensitivity of the signal value to the magnetic field inhomogeneity stems from diffusion, this suppresses the influence of any misrepresentation of the precise gradient waveform, appearing as a tapering-off of the error with increasing frequency in spite of the worsening approximation of the MCF scheme: As long as the MCF time discretization scale is chosen safely below the thermalization time $\Omega^{-1}$, the expected increase of error with increasing frequency due to time-discretization is tamed.

The random walk simulations offered yet another test for the validity
of our result for the oscillating gradient waveforms.  For these
simulations, the suppression of errors with increasing frequency has
to do with statistics. At low frequencies, the expected value of the
average magnetization is zero, implying that the spin-bearers have a
wide distribution of accumulated phase angles, in order to have
resulted in a lot of mutual cancellation. On the other hand, the
expected magnetization is maximized at high frequencies, implying a
sharp distribution of phase angles. In the extreme case of an
infinitely sharp distribution, one can imagine that even a single
walker's accumulated phase angle will almost certainly match the
expected value of the distribution.  Therefore the lower frequency
simulations (with lower signal value) require more realizations
(walkers, trajectories) to sample the phase distribution
adequately. Conversely, at a fixed number of realizations, higher
frequency simulations (with higher signal values) have better
sampling and hence better likelihood of small error.

\subsection{Characterization of anisotropy}

Incorporating a tensorial force constant (while keeping the diffusivity a scalar) leads to a characterization of anisotropy alternative to the commonly employed diffusion tensor model. Here, we shall focus on the explicit results for the Stejskal-Tanner sequence, \ie., Eqs.~\eqref{eq:E.1PFG} and \eqref{eq:Amatrix}. The corresponding expression for the diffusion tensor model is \cite{Stejskal65,Basser94}
\begin{align}
E=\exp \left[ - \gamma^2 \delta^2 \left(\Delta - \frac{\delta}{3} \right) \bm G^\mT \bm D \bm G \right] \label{eq:E_DTI} \ ,
\end{align}
where $\bm D$ is the diffusion tensor. 

\begin{figure}[t] 
	\begin{center}
		\includegraphics[width=8.4cm]{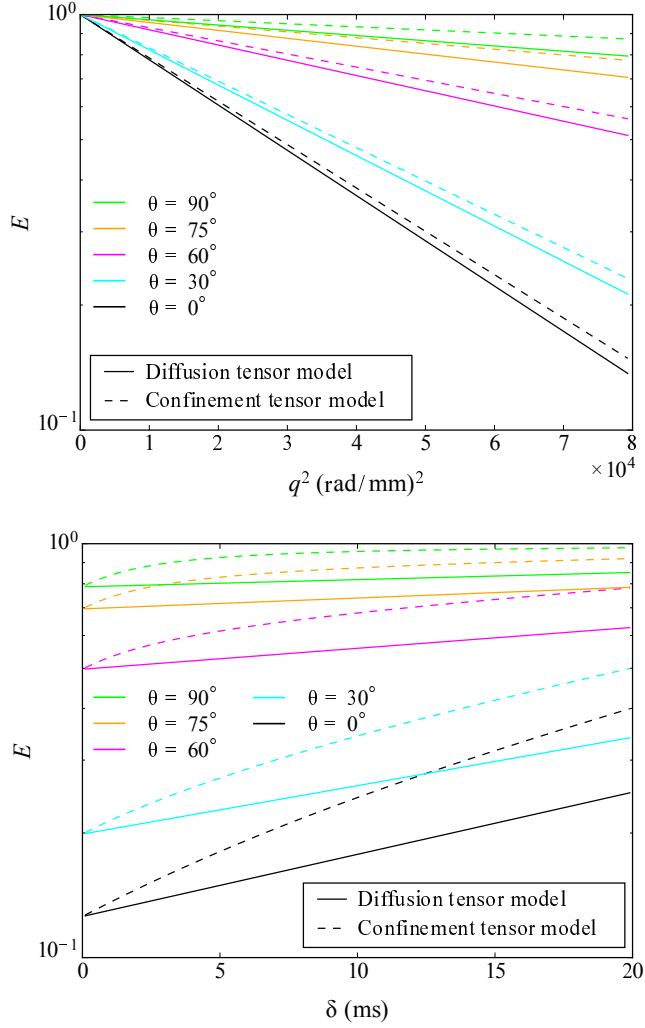}
		\caption{(Color online) The predicted MR signal plotted against the pulse duration in a Stejskal-Tanner pulse sequence. Different curves represent different values of the angle ($\theta$) between the gradient vector and the eigenvector of the confinement tensor associated with its smallest eigenvalue.} \label{fig:StejskalTanner_anisotropy}
	\end{center}
\end{figure}

The diffusivity and temperature were as before while the confinement tensor was taken to be diagonal with eigenvalues chosen such that $C_x=C_y$ corresponded to a pore size of approximately $6 \, \mu$m while $C_z$ was taken to be one-tenth of $C_x$ so that the corresponding pore size along the $z$ axis was roughly $18 \, \mu$m. We chose the eigenvalues of the diffusion tensor through the expression
\begin{align}
D_i = \frac{1}{C_i \Delta} \left( 1- \ee^{-C_i \Delta \DD } \right) \ ,
\end{align}
which was obtained by setting the mean squared displacements, $\langle (\Delta x)^2 \rangle$, implied by the two models equal to each other (see Eq. \eqref{eq:Leff}). Different colors in Figure \ref{fig:StejskalTanner_anisotropy} represent different values for the angle between the gradient direction and the ``fiber,'' \ie., the $z$-axis. 

In the first panel of Figure \ref{fig:StejskalTanner_anisotropy}, we plot the predicted NMR signal against $q^2$. The timing parameters were: $\Delta = 20\,$ms and $\delta=2\,$ms. As expected, the $q^2$ dependence of the signal appears linear for both models on a semi-logarithmic plot. The two models yield essentially the same results; the slight deviation is due to the pulse duration. 

The difference of the models becomes apparent when one considers the dependence of the signal on the timing parameters of the sequence. To illustrate this, we depict the $E$ vs. $\delta$ curves in the second panel of Figure \ref{fig:StejskalTanner_anisotropy}. In these plots, the $q$ values were fixed to $90\pi \,$rad/mm and the pulse separation was again $\Delta = 20\,$ms. In this case, the diffusion tensor model implies a linear appearance in the semi-logarithmic plots. However, the dependence in the confinement tensor model is more complicated with non-linearity becoming quite substantial in the most restricted directions.

\section{Discussion and Conclusion}

Parabolic potentials are useful in numerous physics problems when one is interested in modeling slight deviations away from the equilibrium state; when the perturbation is small, any potential can be approximated by a quadratic function. In this study, however, the utility of employing a parabolic potential model was due to a different reason. Following previous work \cite{Stejskal65,LeDoussal92,MitraHalperin95}, our development can be regarded to yield an alternative model of restricted diffusion, which was the main motivation of this study. Its resemblance to restricted diffusion is due to the time-dependence of the underlying mean square displacements (hence the diffusion coefficients). The mean square displacement implied by the propagator in \eqref{eq:propagator} is $\langle (\Delta x)^2 \rangle = 2 C^{-1}(1-e^{-D_0 C t})$, which approaches a plateau as $t\rightarrow \infty$ as expected in restricted geometries. 

Despite this resemblance, the signal implied by the parabolic potential model has an important difference from that for restricted diffusion models. The phases of the random walkers turn out to be Gaussian-distributed \cite{LeDoussal92}. Thus, the logarithm of the predicted signal for each isolated pore has only a quadratic term in gradient magnitude. Consequently, features that are visible only at large diffusion sensitivities (\eg., diffraction-like effects \cite{Callaghan91,Ozarslan_JMR07}) are not accomodated by the parabolic potential model. Although this seems like a limitation of the model, for most acquisitions involving microscopic pores and weak gradients, the low diffusion sensitivity behavior is the only regime of the signal that can be measured. Under these conditions, the observed higher order terms emerge predominantly from heterogeneities within the sample rather than true compartmental effects. Heterogeneity-induced effects would be captured by the present model as well if the signal is represented as the superposition of signals associated with a distribution of confinement values, \ie., potentials, similar to what is done in multiple diffusion tensor \cite{Inglis01} and diffusion tensor distribution \cite{Jian07,Westin16NI} representations. 

\begin{figure}[!t] 
	\begin{center}
	\includegraphics[width=8.4cm]{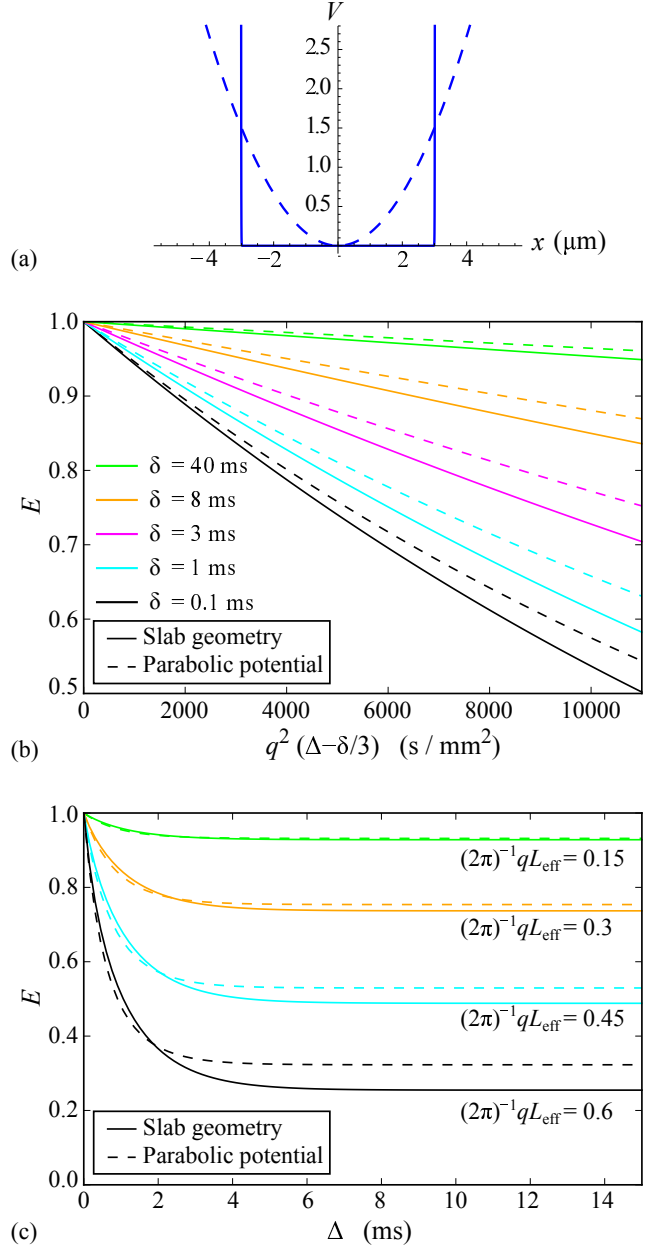}
		\caption{(Color online) (a) Continuous line: Profile of the dimensionless potential due to restricted diffusion taking place between two parallel plates separated by $6 \,\mu$m. Dashed curve:  Parabolic potential that could be employed for the same slab geometry. (b) The signal for the traditional Stejskal-Tanner sequence plotted against the $b$-value for $\Delta = 50 \, \mathrm{ms}$. (c) The signal plotted against the pulse separation for the Stejskal-Tanner sequence. } \label{fig:compare2slab}
	\end{center}
\end{figure}

To illustrate the above points and better assess the resemblance of the Hookean model to restricted diffusion, we consider molecules diffusing between two infinite parallel plates separated by $6\, \mu$m. For this geometry, the potential is zero within the slab and infinite at the walls. Figure \ref{fig:compare2slab}a depicts the potential profile for the slab geometry (continuous line) as well as the parabolic potential (dashed line) that could be used in lieu of it. For the latter, the confinement value was obtained through \eqref{eq:Leff} as 0.33 $\mu$m$^{-2}$. In Figure \ref{fig:compare2slab}b, we plot the signal against the quantity $q^2 (\Delta - \delta /3)$, which is frequently referred to as the ``b-value,'' and fully characterizes the signal decay for free diffusion \cite{Ozarslan15cmra}. Here, curves in different colors represent different values of the pulse duration. For each pulse duration, the $b$-value was varied by changing the gradient magnitude. Having different signal values for the same $b$-value is a further manifestation of the failure of the free diffusion model in characterizing restricted diffusion. Employing the parabolic potential however, yields a behavior similar to that observed for restricted diffusion. We observe that the deviation between the signal values for restricted and Hookean models becomes more pronounced as the $b$-value is increased. We show the dependence of the signal on the separation of pulses in Figure \ref{fig:compare2slab}c. To be able to simulate shorter diffusion times, the pulse duration was fixed at $1\, \mu$s. Here, different colors represent different values of the wavenumber $q$. As expected, the signals for both problems reach constant values as the pulse separation (diffusion time) is prolonged. The curves for the two models are nearly indistinguishable at small attenuations. For stronger diffusion sensitivity, the transition to the long diffusion-time regime occurs slightly more quickly in the case of parabolic potential, yielding a larger eventual signal. We reiterate that the comparisons here are based on the association in \eqref{eq:Leff}, which is only one way of assigning a pore size to a particular confinement value. Many more associations could be made (\eg., based on zero displacement probabilities) that would change the discrepancies between the curves in Figure \ref{fig:compare2slab} though the qualitative resemblance afforded by the parabolic potential model would prevail.

The first approach we employed to tackle the problem involved posing the NMR signal as a path integral and evaluating it explicitly. As exemplified by the cases of Stejskal-Tanner and oscillating gradient sequences, this approach can be employed to obtain analytical results for the NMR signal featuring arbitrary gradient waveforms. Formerly, the path integral method was used to derive the expression describing the influence of free diffusion \cite{Ozarslan15cmra}. In an earlier publication, the quadratic term of the signal decay due to restricted diffusion was calculated via an explicit evaluation of the path integral for the simple geometries of parallel plates, cylinders, and spheres \cite{OzarslanJCP08}. Employing the former two geometries in perpendicular directions, the solution for a capped cylinder geometry can be obtained \cite{Ozarslan_JMR09}, which had been used to model restricted diffusion in an anisotropic pore. However, because the capped cylinder is invariant to rotations about its axis, this model has only two size parameters---its radius and length, thus limiting its applicability. On the other hand, our model has three unique confinement values ($C_i$, where $i=1,2,3$), each of which can be related to an effective pore size along three spatial directions. In this respect, our model is more general than the capped cylinder geometry for modeling anisotropy. 

We note that an alternative geometry involving three size parameters is the rectangular prism. The MR signal influenced by diffusion within a rectangular prism could be obtained by employing the solution for parallel plates in three orthogonal directions. However, even when the three size parameters are set equal, the resulting geometry is not truly isotropic, making it difficult to isolate the influence of structural anisotropy. This limitation, which is present also for the capped cylinder geometry, is overcome by our method. Setting all three principal confinement values equal to each other provides the solution for a perfectly isotropic potential akin to the case of diffusion taking place within a sphere. On the other hand, setting only two of the three eigenvalues of the confinement tensor equal to each other provides the solution for diffusion within a cylinder-like pore.

As exemplified in Figure \ref{fig:StejskalTanner_anisotropy}, compared to the diffusion tensor model, which is based on the assumption of free diffusion, our model has a different dependence on the timing parameters of the gradient waveforms. This issue is likely to be important when sophisticated pulse sequences are employed. Indeed, several different pulse sequences have been studied in recent years employing pulsed \cite{Cory90b,Mitra95,Ozarslan_JMR07,Paulsen_inpress} and continuous \cite{Eriksson13,Westin14miccai,Ozarslan_ISMRM14_RFG} waveforms. Especially in the context of modeling local diffusion anisotropy, one is faced with the choice between the Gaussian diffusion tensor model \cite{Lasic14,Westin16NI}, and restricted capped cylinder model \cite{Ozarslan_JMR09,Benjamini14shape}. Although, the restricted character of the model is desirable, the capped cylinder model had two limitations, which may have prompted some studies to adopt the diffusion tensor model instead: (i) it has only two size parameters, so is more limited than the diffusion tensor model, (ii) the solutions are rather complicated. We believe the confinement tensor model introduced in this study overcomes both of these limitations making it suitable to characterize global (macroscopic) as well as local (microscopic) diffusion in heterogeneous media. For example, ``diffusion imaging with confinement tensor (DICT)'' could be performed by estimating the confinement tensor via Eqs. \eqref{eq:E.1PFG} and \eqref{eq:Amatrix} for each voxel of a data set comprising multi-directional diffusion-weighted images, much like Eq. \eqref{eq:E_DTI} is employed in diffusion tensor imaging (DTI) studies.  

Although our formulation has focused for the most part on parabolic potentials and linear gradients, the MCF and random walk simulations can be generalized to more general cases. For example, spatially nonlinear gradients can be incorporated with ease by simple manipulations of the operators employed in the MCF technique \cite{Laun12jcp}. Other potentials can also be incorporated by employing exact or approximate solutions of the corresponding Helmholtz problem. Finally, the work presented in this article may be relevant in very different contexts. For example, our approach here could be expanded to offer some utility in modeling noise induced by Brownian motion in magnetic resonance force microscopy \cite{Sidles91,Sidles93}.

In conclusion, we have presented an in-depth study on the effect of diffusion taking place in the presence of a parabolic potential on the NMR signal. The scheme enables one to incorporate restriction-induced effects when general gradient waveforms are employed. The study lends itself to a new formulation for diffusion anisotropy featuring three parameters like the diffusion tensor model. However, by overcoming the limitations due to the assumption of free diffusion, the confinement tensor model could be considered a powerful alternative to the diffusion tensor model for characterizing anisotropy.

\section*{Acknowledgments}
We thank Dr. Chunlei Liu for his comments on the nomenclature. The authors acknowledge funding from T\"UB\.ITAK-EU COFUND Project 114C015, Bo\u gazi\c ci University Research Fund (grant number 8521), Bilim Akademisi BAGEP 2014 Award, NIH R01MH074794, and Swedish Foundation for Strategic Research (No. AM13-0090).

\appendix

\section {Derivations within the path integral formalism} \label{app:pathint}

This appendix is reserved for the details of the derivation involved
in arriving at the analytical expression \eqref{eq:E.3D} for the
average magnetization, or NMR signal. We will begin with the
discretization of the path integral \eqref{eq:path.int}, followed by
its evaluation (and continuum limit) for a parabolic potential in one
spatial dimension, and then generalize to higher dimensions.

\subsection{Discretization}

As is usual, the path integral is evaluated via slicing the time interval $0<t<\te$ into a number $n$ of time steps, and before the continuum limit ($n \rightarrow \infty$) is considered, discretized versions of the integration measure and the integrand are employed. For what follows, we define the small time step $\tau = \te/n$, and $t$ is replaced by $j \tau$ for the discretization, with $j$ an integer in the interval $[0,n]$. Also, time integrals $\int \dd t$ become $\tau \sum_j$. Following the definition of the variables
$\bfr_j = \bfr( j \tau) $ and $ \bm{q}_j = \gamma \tau \bm{G} ( j \tau)$, the discretization amounts simply to
\begin{subequations} \label{eq:discretizn}
	\begin{align}
	\gamma \int_0^{\te} \dd t \, \bm{G} (t) \cdot \bfr (t)
	&\rightarrow \sum_{j=0}^n \bm{q}_j \cdot \bfr_j\;, \\
	\mathscr{D} \bfr (t) &\rightarrow \prod_{j=0}^n \dd \bfr_j \;, \\
	\mathscr{P} [\bfr(t)] &\rightarrow \rho (\bfr_0) \prod_{j=1}^n 
	P_\tau (\bfr_{j-1}, \bfr_j) \;,
	\end{align}
\end{subequations} 
where $\rho (\bfr_0)$ is the initial probability distribution of spins, and $P_t (\bfr', \bfr)$ is the probability of a particle ending up at the \emph{later} position $\bfr$ given that it was at the \emph{earlier} position $\bfr'$ a time $t$ ago.\footnote{We have also committed a slight abuse of notation by denoting volume integrals as $\dd \bfr$.} As a result, the time-sliced path integral becomes
\begin{align}
E_{\tau} = \int \dd \bfr_0 \rho (\bfr_0) \ee^{- \im \bm{q}_0 \cdot \bfr_0}
\prod_{j=1}^n \dd \bfr_j P_{\tau} (\bfr_{j-1}, \bfr_j) \ee^{-\im
	\bm{q}_j \cdot\bfr_j} \;, \label{eq:E.discrete.1D}
\end{align}
where the subscript $\tau$ reminds us that this is not yet the exact
path integral, but a discretized approximation. Also, note how the
integrals over later positions are nested inside those of the earlier
ones.

The above discussion holds for any (time-independent) potential and
number of dimensions. However, evaluating the (discretized) path
integral \eqref{eq:E.discrete.1D} requires an explicit form for the
\emph{propagator} $P_\tau (\bfr', \bfr)$. We focus first on the one
dimensional case with a parabolic potential.

\subsection{Hookean force in one dimension}  \label{sec:analytical1D}


The propagator $P_t (x', x)$ for diffusion under the effect of a dimensionless potential $V(x) = (1/2) C x^2$ has the form \cite{Smoluchowski13,Uhlenbeck30}
\begin{align} 
P_t (x', x) = \frac {1} {\sqrt{2 \pi \sigma_t^2}}
\exp \!\left[ -\frac {(x - s_t x')^2} {2 \sigma_t^2} \right]
\label{eq:propagator} \ .
\end{align}
With $\DD$ being the bare diffusion constant, the inverse time $\Omega$ is defined as
\begin{subequations} \label{eq:OSS}
	\begin{align}
	\Omega = \DD C \;, \label{eq:Omega}
	\end{align}
	the shorthand symbols $s_t$ and $\sigma_t$, having been used above, can be expressed as
	\begin{align}
	s_t = \ee^{-\Omega t} \ , \label{eq:s}
	\end{align}
	and
	\begin{align}
	\sigma_t^2 = \left(1-s_t^2 \right) C^{-1}
	\;. \label{eq:sigma^2}
	\end{align}
\end{subequations}

Note that the propagator~\eqref{eq:propagator} is a normalized
Gaussian in its later argument $x$, centered at a point that depends
on its earlier argument $x'$. Despite its Gaussian appearance, this
propagator has an important characteristic that is consistent with
restricted diffusion. To illustrate this point, we shall consider the
long time limit ($t \rightarrow \infty$) of the propagator. In this
regime, the dependence on the initial position disappears as expected
for restricted diffusion. It is instructive to also consider the net
displacements, \ie., $\Delta x = x - x'$. Let $\rho(x')$ denote the
equilibrium spin density. The second moment of net displacements,
defined through the expression
\begin{align}
\langle (\Delta x)^2 \rangle = \int \dd x' \, \rho(x') \int \dd\Delta x \, (\Delta x)^2 \, P_t(x',x'+\Delta x)  \label{eq:MSD}
\end{align}
diverges for free diffusion as the diffusion time is prolonged. However, the same quantity for the propagator in Eq. \eqref{eq:propagator} asymptotically approaches the constant value of $2/C$. On the other hand, if one considers a restricted diffusion process taking place between two infinite plates separated by a distance $L$, this quantity similarly approaches a constant value, which is $L^2 / 6$. Setting the latter two quantities equal to each other, one can obtain an effective pore size given by 
\begin{align} 
L_{\mathrm{eff}} = \sqrt{\frac{12 D_0}{\Omega}} = \sqrt{\frac{12}{C}}
\label{eq:Leff} \ .
\end{align}
We refer to this relationship in the Results section.

\subsubsection{Evaluation of the signal}

The form of the propagator~\eqref{eq:propagator} helps a great deal in evaluating the signal analytically. Indeed, the discretized path integral~\eqref{eq:E.discrete.1D} involves only familiar integrals, but each integration is affected by the result of that nested immediately inside it. Consider, for instance, the inner-most integral (over $x_n$):
\begin{align}
\int \dd x_n \, P_\tau (x_{n-1}, x_n) \ee^{-\im q_n x_n} = 
\, \ee^{-\half \sigma_\tau^2 q_n^2} \ee^{-\im s_\tau q_n x_{n-1}} \;.
\end{align}
The first factor has nothing to do with the remaining integrals in \Eqref{eq:E.discrete.1D}. The second factor, on the other hand, combines with the exponential in the next integral (over $x_{n-1}$), shifting its ``wave number'' $q_{n-1}$ by an amount $s_\tau q_n$. One
can see that, with later (large index) wave numbers leaking into the phase factors of earlier (smaller index) integrations as such, and getting multiplied by $s_\tau$ at each step, it should be useful to define
\begin{align}
Q_j = \sum_{i=j}^n s_\tau^{i-j} q_i \ . \label{eq:Qj}
\end{align}
(The index $i$ is not to be confused with the imaginary unit $\im$.) Following this recursion just described, one finds
\begin{align}
E_\tau = \exp \!\left( - \frac {\sigma_\tau^2}{2} \sum_{j=1}^n
Q_j^2 \right) \int \dd x_0 \, \rho (x_0) e^{- \im Q_0 x_0}
\;. \label{eq:lastint}
\end{align}
We assume that the distribution of spin-carriers has had enough time to reach equilibrium before the gradient sequence is applied. Hence, the relation $\rho (x_0) = P_\infty (x_{-1}, x_0)$ can be employed (Eqs.\ \eqref{eq:propagator} -- \eqref{eq:OSS} in the limit $t
\rightarrow \infty$), with $x_{-1}$ being some arbitrary pre-initial position. We can thus perform the last remaining integration similarly to the previous ones to find
\begin{align}
E_\tau =& \exp \!\left( - \frac {\sigma_\tau^2}{2} \sum_{j=1}^n
Q_j^2 - \frac{ \sigma_\infty^2} {2} Q_0^2 \right) \nonumber \\
=& \exp \!\left( - \frac{\DD} {2 \Omega}\left( 1- e^{-2 \Omega \tau} 
\right) \sum_{j=1}^n Q_j^2 - \frac {\DD} {2 \Omega} Q_0^2 
\right) \ . \label{eq:E.discrete}
\end{align}

\paragraph*{Continuum limit.}

In the continuum limit where $\tau = \te/n \rightarrow 0$ as $n \rightarrow \infty$, we have $j \tau \rightarrow t$ and $\tau \sum_j \rightarrow \int \dd t$. Recalling moreover that $Q_j$ really stands for $Q(j \tau)$, the $\tau \rightarrow 0$ limit of the average
magnetization \eqref{eq:E.discrete} is easily obtained as
\begin{align}
E = \exp \! \left( - \DD \int_0^{\te} \dd t \, Q^2 (t) -\frac {\DD}
{2 \Omega} Q^2(0) \right) \ . \label{eq:E.cont}
\end{align}
The quantity $Q(t)$ follows from \Eqref{eq:Qj} similarly. Recalling that $q_j = \gamma\tau G (j \tau)$, and using \Eqref{eq:s}, the expression \eqref{eq:Qj} for $Q_j$ can be recast as
\begin{align}
Q_{j} = \gamma  \sum_{i=j}^n \tau G(i\tau) \ee^{-\Omega \tau(i-j)}
\;.
\end{align}
In the $\tau \rightarrow 0$ limit, one finds
\begin{align}
Q(t) = \gamma \int_t^{\te} \dd t' \, 
\ee^{- \Omega (t'-t)} G(t') \ . \label{eq:Q(t)}
\end{align}
Eqs.\ \eqref{eq:E.cont} and \eqref{eq:Q(t)} yield the final result for the NMR signal for the case of molecules diffusing under the influence of a Hookean restoring force in one dimension.

\paragraph*{Free diffusion.}
Here, we show that our expression \eqref{eq:E.cont} for the signal agrees with previous results for free diffusion NMR signal \cite{Karlicek80} 
\begin{align}
E_{\rm free} = \exp \! \left[ -\DD \gamma^2 \int_0^\te \dd t\,
\left( \int_0^t \dd t'\, G(t') \right)^2 \right] \label{eq:E.free}
\end{align}
in the limit of vanishing potential ($f \rightarrow 0$). In this limit, $\Omega \rightarrow 0$, and thus the exponential in \Eqref{eq:Q(t)} drops. Along with the gradient echo condition, $\int_0^\te \dd t' \, G (t') =0$, we have
\begin{align}
Q_{\rm free} (t) = - \gamma \int_0^t \dd t' \, G(t') \;, \label{eq:Q.free}
\end{align}
and $Q(0) = 0$. Therefore, the first term in our expression for the signal \eqref{eq:E.cont} matches the free diffusion NMR signal \eqref{eq:E.free}. The second term in \Eqref{eq:E.cont} does not seem to disappear, due to the $\Omega^{-1} \rightarrow \infty$
prefactor. However, if one recalls that this term is simply the remaining integral in \Eqref{eq:lastint}, then it is recognized that the integral becomes $\int \dd x \, \rho (x)$ when $Q_0 = Q(0)=0$, which is unity by definition.

\subsection{Hookean force in three dimensions} \label{sec:3Dpath}


In three dimensions, the most general harmonic potential (assuming the attraction center is at the origin) is of the form
\begin{align}
V = \half \bfr^{\mT} \bm{C} \bfr \ ,
\end{align}
where the confinement tensor $\bm{C}$ is real and can be assumed to be symmetric without loss of generality. There exists, then, a rotation matrix $\rotb$ such that
$\rotb \rotb^\mT= \bm{1}$, and
\begin{align}
V = \half \tilde{\bm{r}}^\mT \tilde{\bm{C}} \tilde{\bm{r}} \ ,
\end{align}
where $\tilde{\bm{C}} = \rotb^{\mT} \bm{C} \rotb$ is diagonal, and $\tilde{\bm{r}} = \rotb^{-1} \bm{r}$. In this basis denoted by tildes, the potential has the coordinates $\tilde{x}$, $\tilde{y}$, and $\tilde{z}$ decoupled, and so does the Smoluchowski equation that describes the propagator $P_t (\tilde{\bfr}', \tilde{\bfr})$. Therefore, the propagator simply factorizes into three instances of the propagator~\eqref{eq:propagator} for each direction:
\begin{align}
P_t (\tilde{\bfr}', \tilde{\bfr}) = \prod_{i=1}^3 \frac {1}
{\sqrt{2 \pi \sigma_{t,i}^2}} \exp \!\left[ -\frac {(\tilde{r}_i -
	s_{t,i} \tilde{r}'_i)^2} {2 \sigma^2_{t,i}} \right]
\ , \label{eq:propagator.3}
\end{align}
where Eqs.\ \eqref{eq:OSS} become
\begin{subequations} \label{eq:OSS.i}
	\begin{align}
	\Omega_i =& \DD C_i \;, \label{eq:Omega.i}\\
	s_{t,i} =& \ee^{-\Omega_i t} \ , \\
	\sigma_{t,i}^2 =& \left(1-s_{t,i}^2 \right) C_i^{-1} \;,
	\end{align}
\end{subequations}
with $C_i$ denoting the $i$th eigenvalue of $\bm{C}$, \etc.

By virtue of this factorization, one sees that the evaluation of the path integral proceeds exactly the same way as before, only this time it is three-fold:
\begin{align}
E =& \prod_{i=1}^3 \exp \! \left( - \DD \int_0^\te \dd t \,
\tilde{Q}_i^2 (t) - \frac {\DD}{2 \Omega_i} \tilde{Q}^2_i (0)
\right) \label{eq:Etilde.3D} 
\end{align}
with
\begin{align}
\tilde{Q}_i (t) = \gamma \int_t^\te \dd t' \,
\ee^{-\Omega_i (t'-t)} \tilde{G}_i (t') \;. \label{eq:Qtilde.3D}
\end{align}
Recall that the two equations above are valid in the rotated basis where the confinement tensor $\bm{C}$ is diagonal (hence the tildes).

In order to revert to the lab frame (the un-rotated basis), note that $\Omega_i^{-1}$ and $\ee^{-\Omega_i (t'-t)}$ are just the $i$th eigenvalues of the matrices $\bm{\Omega}^{-1}$ and $\ee^{- \bm{\Omega} (t'-t)}$ which are diagonalized by the same rotation $\rotb$ as $\bm{\Omega}$, and hence $\bm{C}$---see \Eqref{eq:Omega.i}. Exploiting as well the rotation properties of $\rotb$, one finds 
\begin{align}
E = \exp \! \left( - \DD \int_0^\te \dd t \, | {\bm{Q}} (t)|^2 - \frac
{\DD}{2} {\bm{Q}}^\mT (0) \bm{\Omega}^{-1} {\bm{Q}} (0) \right)
, \label{eq:E.3D.A}
\end{align}
where
\begin{align}
\bm{Q} (t) = \gamma \int_t^\te \dd t' \, \ee^{-\bm{\Omega} (t'-t)} 
\bm{G} (t') \ . \label{eq:Q.3D.A}
\end{align}
These are Eqs.\ \eqref{eq:E.3D} and \eqref{eq:Q.3D} of the main
text. Since one in general does not know what the principal directions
are without previous knowledge about the specimen, these general
expressions are more relevant than the diagonalized versions
\eqref{eq:Etilde.3D} and \eqref{eq:Qtilde.3D}. However, note that in
evaluating the integral in \Eqref{eq:Q.3D} with a given gradient
waveform $\bm{G} (t)$, one may have to make sense of the exponentiated
matrix in the integrand by going back and forth between the lab frame
and the eigen-basis.

\section {MCF formalism} \label{app:MCF}

This appendix is reserved for the technicalities of the MCF
calculations of Section \ref{sec:MCF}.

It is illuminating to consider the BTS equation
\eqref{eq:BTS} as the projection of an operator equation for the
vector $\ket{\MM (t)}$ onto the (dual) vector $\bra{\bfr}$. If one
defines the operators $\LL$ and $\GG (t)$ such that
\begin{align}
&\bra{\bfr} \LL \ket{\MM(t)} = \DD \nabla \cdot \left(
\ee^{- V(\bfr)} \nabla \ee^{ V(\bfr)} \MM (\bfr, t)
\right) \;, \label{eq:LL} \\
&\bra{\bfr} \GG (t) \ket{\MM(t)} = - \im \gamma \bm{G} (t) \cdot \bfr 
\, \MM (\bfr, t) \;, \label{eq:GG}
\end{align}
\Eqref{eq:BTS} may be rewriten, without the projection onto spatial
coordinates, as
\begin{align}
\frac {\del} {\del t} \ket{\MM (t)} = \left[ \LL +
\GG (t) \right] \ket{\MM (t)} \;. \label{eq:BT.sym}
\end{align}
We will refer to the operator $\LL$ defined in \Eqref{eq:LL} as the Smoluchowski operator. 

In the MCF framework, the total time interval is sliced into smaller
successive intervals $t_{j-1} < t < t_{j}$ with $j=1,2, \ldots,
n$. Within each of these intervals, the gradient waveform $\bm{G}
(t)$, hence the operator $\GG (t)$, is considered to be constant. If
the operator $\GG (t)$ has the constant value $\GG_j$ in the interval
$t_{j-1} < t < t_j$, that is,
\begin{align}
\GG (t) =& \GG_j  \quad (t_{j-1} < t < t_j) \;,
\end{align}
then
\Eqref{eq:BT.sym} implies that
\begin{align}
\ket{\MM (t_j)} = \ee^{(\LL + \GG_j) \Delta t_{j}}
\ket{\MM (t_{j-1})}\;,
\end{align}
where $\Delta t_j$ is the duration of the $j$th time interval. Thus,
when all $n$ intervals have elapsed,
\begin{align}
\ket{\MM(t_n)} = \sideset{}{'} \prod_{j=1}^n \ee^{( \LL +
	\GG_j) \Delta t_j} \ket{\MM (0)} \;,
\end{align}
where the prime on the product symbol is to remind us that subsequent
propagators multiply from the \emph{left}.  Finally, the signal
becomes
\begin{align}
E (t) = \int \dd \bfr \, \bra{\bfr} \sideset{}{'} \prod_{j=1}^n
\ee^{( \LL + \GG_j) \Delta t_j} \ket{\MM (0)}
\;. \label{eq:signal.sym}
\end{align}

This is the essence of the MCF formalism. What remains are details of
implementation and evaluation, which vary according to the
specifications of the problem at hand. The expression derived above
becomes increasingly accurate for an arbitrary gradient waveform
$\bm{G}(t)$ with finer time-slicing. Or, it may be considered exact if
the gradient waveform is as a matter of fact constant over finite time
intervals.

\subsection {Eigenfunction expansion}
The first step toward the evaluation of \Eqref{eq:signal.sym} is to
make the expression concrete by choosing a basis to project the
operators and vectors onto.  The two most obvious choices are the
eigenvectors of the operators $\LL$ and $\GG(t)$.  The eigenvectors of
$\GG(t)$, according to \Eqref{eq:GG}, are the same set $\left\{
\ket{\bfr} \right\}$ of vectors as the eigenvectors of the position
operator. Using this basis will result in a path integral type
implementation \cite{Yolcu15Dagstuhl}.

Alternatively, one projects onto the eigenvectors of the Smoluchowski
operator $\LL$, but there is a catch: the operator $\LL$ is not self-adjoint, thus a spectral decomposition may not be available. However, it is similar to a self-adjoint operator in the sense that the operator $\bm{H}=\ee^{\bm{V}/2} \LL \ee^{-\bm{V}/2}$ satisfies $\bm{H}^\dagger=\bm{H}$.\footnote{Here, $\bm{V}$ is the potential operator, such that $\bra{\bm{r}} \bm{V}= V(\bm{r}) \bra{\bm{r}}$. If one considers the self-adjoint wave vector (or momentum) operator $\bm{K}$ such that $\bra{ \bm{r}} \bm{K} = - \im \nabla \bra{\bm{r}}$, the operator $\LL$ given in \Eqref{eq:LL} can be written as $\LL = -\DD \bm{K} \cdot \ee^{-\bm{V}} \bm{K} \ee^{\bm{V}}$. One can verify directly that $\bm{H}=\ee^{\bm{V}/2} \LL \ee^{-\bm{V}/2}=-\DD \ee^{\bm{V}/2} \bm{K} \cdot \ee^{-\bm{V}} \bm{K} \ee^{\bm{V}/2}$ satisfies $\bm{H}^\dagger = \bm{H}$.} This allows $\LL$ to have a real spectrum and a complete set of bi-orthogonal eigenvectors, as we will briefly discuss below.

If the self-adjoint operator $\bm{H}$ has a set of real eigenvalues $\lambda_k$ and corresponding eigenvectors $\ket{\tilde{u}_k}$ satisfying
\begin{align}
\bm{H} \ket{\tilde{u}_k} = \lambda_k \ket{\tilde{u}_k} \;, 
\end{align}
it can be verified via straightforward algebraic manipulations that the vector $\ket{u_k}=\ee^{- \bm{V}/2} \ket{\tilde{u}_k}$ satisfies
\begin{align}
\LL \ket{u_k} = \lambda_k \ket{u_k} \;,
\end{align}
\ie., $\ket{u_k}$ is an eigenvector of $\LL$ with eigenvalue $\lambda_k$.
Similarly, the vector
\begin{align}
\ket{w_k} = \ee^{\bm{V}} \ket{u_k} \;, \label{eq:wu}
\end{align}
satisfies the eigenvalue equation
\begin{align}
\bra{w_k} \LL = \lambda_k \bra{w_k} \;.
\end{align}
It is these left and right eigenvectors, taken in pairs, which satisfy relations of orthonormality
\begin{align}
\braket{w_k}{u_l} = \delta_{kl} \label{eq:orthuw}
\end{align}
and completeness
\begin{align}
\bm{1} = \sum_k \ket{u_k} \bra{w_k} \;, \label{eq:compuw}
\end{align}
as a result of the orthonormality and completeness of the eigenvectors $\ket{\tilde{u}_k}$ of the self-adjoint operator $\bm{H}$. 

Now, we can use the completeness relation \eqref{eq:compuw} to rewrite
the NMR signal \eqref{eq:signal.sym} as
\begin{align}
E = \sum_{k,\ell} \int \dd \bfr \, \braket{\bfr}{u_k} \bra{w_k} \sideset{}{'} \prod_{j=1}^n
\ee^{( \LL + \GG_j) \Delta t_j} \ket{u_\ell} \braket{w_\ell} {\MM (0)}
\;. \label{eq:signal.comp}
\end{align}

\subsubsection{The equilibrium state}

When no perturbation via the gradient $\bm{G} (t)$ is applied, the BTS
equation \eqref{eq:BTS} must lead to the equilibrium magnetization density $\sim
\ee^{- V(\bfr)}$ in the $t\rightarrow \infty$ limit. Such
stability of solutions requires the eigenvalues $\lambda_k$ to be
bounded from above: $\lambda_k \le 0$.  The eigenfunction
corresponding to the maximal eigenvalue $\lambda_0=0$, therefore, is
(proportional to) the equilibrium ($t \rightarrow \infty$) solution,
\begin{align}
u_0 (\bfr) = Z^{-1/2} \ee^{-V(\bfr)} \;,  
\end{align}
with $Z$ being a normalization factor given as
\begin{align}
Z = \int \dd \bfr \, \ee^{-V(\bfr)} \; .
\end{align}
Also note that, according to \Eqref{eq:wu}, we have
\begin{align}
w_0 (\bfr) = Z^{-1/2} \;, \label{eq:w0}
\end{align}
for the dual eigenfunction corresponding to the equilibrium
eigenvalue $\lambda_0 = 0$.

We can exploit the properties mentioned above of the equilibrium
eigenfunctions to simplify \Eqref{eq:signal.comp}. First, if the
initial magnetization density is at equilibrium, then $\ket{\MM (0)} =
Z^{-1/2} \ket{u_0}$, meaning $\braket{w_\ell}{\MM (0)} = Z^{-1/2}
\delta_{\ell 0}$. Second, we can (ab)use \Eqref{eq:w0} to the effect,
\begin{align}
\int \dd \bfr \, \braket{\bfr} {u_k} = Z^{1/2} \int \dd \bfr \, \braket
{w_0} {\bfr} \braket{\bfr} {u_k} = Z^{1/2} \delta_{0k} \; .
\end{align}
These two manipulations eliminate both summations, and the integral in
\Eqref{eq:signal.comp}, to yield
\begin{align}
E 
&= \bra{w_0} \sideset{}{'} \prod_{j=1}^n
\ee^{( \LL + \GG_j) \Delta t_j} \ket{u_0} \;. \label{eq:signal.exp}
\end{align}

\subsubsection{Matrix elements}

The above analysis shows that the signal can be written as the top matrix element of the propagator, expressed in the eigenbasis (and the dual basis) of the Smoluchowski operator. While it may be possible to evaluate this matrix element analytically, it is usually necessary to resort to numerical computations, both to compute the product, and also the exponentials. To this end, one uses the matrix elements
\begin{align}
\Lambda_{k\ell} &= \bra {w_k} \LL \ket{u_\ell} =
\lambda_\ell \delta_{k\ell} \quad \text{(no sum implied)}
\;, \label{eq:Lambdakl} 
\end{align}
and, using \Eqref{eq:GG}, 
\begin{align}
\Gamma_{k\ell} (t) = \bra{w_k} \GG(t) \ket{u_\ell} = - \im \gamma \bm{G}
(t) \cdot \bra{w_k} \bm{R} \ket{u_\ell} \;, \label{eq:Gammakl}
\end{align}
where the latter matrix element of the position operator may be
computed, for instance, through
\begin{align}
\bra{w_k} \bm{R} \ket{u_\ell} = \int \dd \bfr \, w_k^*(\bfr) \bfr
u_\ell (\bfr) \; . \label{eq:Rkl}
\end{align}

This is most of what we can say without specifying the potential
$V(\bfr)$ and the number of dimensions.

\subsection{The case of parabolic potential}

\subsubsection {MCF for the one-dimensional problem}

Now, we will restrict to one spatial dimension, and the dimensionless potential
$V(x)=(1/2) C x^2$. In this case, the BTS equation reads
\begin{align}
\frac {\del \MM} {\del t} &= \DD \frac {\del^2 \MM} {\del x^2} + C \DD \frac {\del (x\, \MM)} {\del x} - \im \gamma G (t) x \, \MM \;. \label{eq:BS.1D}
\end{align}
The eigenvalues and eigenfunctions of the Smoluchowski operator $\LL$
can be found from 
\begin{align}
\DD \frac {\dd^2 u_k} {\dd x^2} + C \DD \frac {\dd (x\, u_k)} {\dd x} = \lambda_k u_k \;,
\end{align}
to be
\begin{align}
\lambda_k = - k C \DD \quad (k=0,1,2, \ldots) \;, 
\end{align}
and
\begin{align}
u_k (x) = \left(\frac {C} {2 \pi}\right)^{\frac{1}{4}} \frac{\ee^{-C x^2/2}} {\sqrt{2^k k!}} H_k \!\left( \sqrt{\frac{C}{2}} x \right) \;, \label{eq:uk.1D}
\end{align}
where $H_k(z)$ is a Hermite polynomial. Recall, also, from
\Eqref{eq:wu} that $w_k(x) = \ee^{V(x)} u_k (x)=\ee^{C x^2/2} u_k (x)$.

Immediately, one can evaluate the matrix element \eqref{eq:Gammakl} as
\begin{align}
\Gamma_{k\ell} (t) &= -\im \gamma G (t) \int \dd x \, w_k (x) x
u_\ell (x) \nonumber \\ &= \frac {- \im \gamma G(t) 2^{\frac{k-\ell}
		{2}} k!}  {\sqrt{C} \left( \frac {k+\ell-1} {2}\right)!
	\left( \frac {\ell+1 -k} {2}\right)! \left( \frac {1+k-\ell}
	{2}\right)!}  \;,
\end{align}
for $\vert k-\ell\vert=1$ ($\Gamma_{k\ell}=0$ otherwise) \cite{Gradshteynbook}. The
matrix element \eqref{eq:Lambdakl}, on the other hand, is simply
\begin{align}
\Lambda_{k\ell} = -k C \DD \delta_{k\ell} \quad \text{(no sum implied)} \;.
\end{align}
One can then employ, for a given gradient waveform $\bm{G}(t)$, a
computer to use these matrix elements for $\LL+\GG (t)$ the
computation of the exponentials and the product in the expression
\eqref{eq:signal.exp} for the NMR signal.

\subsubsection{MCF for the three-dimensional problem}

In the case of a three dimensional anisotropic potential, written as $V(\bfr)=
(1/2) r_i C_{ij} r_j$ in index notation, the BTS equation becomes
\begin{align}
\frac {\del \MM} {\del t} &= \DD \del_i \del_i \MM + C_{ij}
\DD \del_i (r_j\, \MM) - \im \gamma G_i (t) r_i \, \MM
\;. \label{eq:BS.3D}
\end{align}
Observing the middle term on the right hand side, one sees that performing a rotation to diagonalize the tensor $\bm{C}$ will decouple the equations over each rotated coordinate ($\tilde{\bm{r}}=\rotb^{-1} \bm{r}$ as in Sec.~\ref{sec:3Dpath}). The eigenfunctions hence factorize,
\begin{align}
u_{\bm{k}} (\tilde{\bm{r}}) = \prod_{i=1}^3 u_{k_i} (\tilde{r}_i) \;,
\label{eq:uk.3D}
\end{align}
with $u_{k_i}(\tilde{r}_i)$ as given in \Eqref{eq:uk.1D}, 
while the eigenvalues of the entire equation become
\begin{align}
\lambda_{\bm{k}} = - \DD C_i k_i \;, \label{eq:lambda.3D}
\end{align}
where $C_i$ are the eigenvalues of the confinement tensor (note the
vector index $\bm{k}=(k_1,k_2,k_3)$ and the sum implied over $i$).
The matrix $\Lambda_{\bm{k} \bm{\ell}}$ is just made up of the
eigenvalues above. 

The NMR signal \eqref{eq:signal.exp} is computed using extensions of Eqs.\ \eqref{eq:Lambdakl}, \eqref{eq:Gammakl}, and \eqref{eq:Rkl}, into three dimensions with rotated coordinates. Noting that $\bm{G}\cdot \bm{r} = \bm{G}^\mT \rotb \rotb^{-1} \bm{r} = \tilde{\bm{G}} \cdot \tilde{\bm{r}}$, we have
\begin{align}
\Lambda_{\bm{k}\bm{\ell}} =& \lambda_{\bm{\ell}} \delta_{\bm{k}\bm{\ell}}
\quad \text{(no sum implied)} \;, \\
\Gamma_{\bm{k}\bm{\ell}} =& -\im \gamma \tilde{\bm{G}}(t) \cdot
\int \dd \tilde{\bm{r}} \, w^*_{\bm{k}} (\tilde{\bm{r}})
\tilde{\bm{r}} u_{\bm{\ell}} (\tilde{\bm{r}}) \;,
\end{align}
where Eqs.\ \eqref{eq:lambda.3D} and \eqref{eq:uk.3D} will be used.

The orientation of the potential (hence the rotation matrix $\rotb$) is not known prior to the measurement of the specimen. Therefore these results are not as general as those of Sec.\ \ref{sec:3Dpath}. The derivation of such general expressions seems intractable to us. However, this does not render the above formulas useless. The ``natural'' frame of reference, hence the matrix $\rotb$, established by the orientation of the specimen can be estimated from the data set, \eg., via employing the diffusion tensor model. The gradient waveform can be transformed into this natural reference frame in which $\bm C$ is diagonal, and the above formulation can be used subsequently \cite{OzarslanNI13}.

\end{document}